\documentclass[aps,prc,superscriptaddress,twoside,twocolumn,nofootinbib,showpacs]{revtex4-2}
\usepackage{amsmath,amssymb}
\usepackage{mathtools}
\usepackage{bbold}
\usepackage[usenames, dvipsnames]{xcolor}
\usepackage{braket}
\usepackage{graphicx}
\usepackage{slashed}
\usepackage{booktabs}
\usepackage{tensor}
\usepackage{multirow}
\usepackage{mwe}
\usepackage{changes}
\usepackage{enumerate}
\usepackage{stackrel}
\usepackage{physics}
\usepackage{bm}
\usepackage{multirow}

\allowdisplaybreaks

\newcommand{\tcb}{\textcolor{blue}}

\usepackage{epsfig}
\usepackage{subfigure}
\usepackage{float}

\allowdisplaybreaks

\begin{document}
\title{Neural Quantum States for Light Nuclei \\ with Chiral Two- and Three-Body Interactions}

\author{Pengsheng Wen}
\email{pswen2019@physics.tamu.edu}
\affiliation{Cyclotron Institute, Texas A\&M University, College Station, TX 77843, USA}
\affiliation{Department of Physics and Astronomy, Texas A\&M University, College Station, TX 77843, USA}

\author{Alexandros Gezerlis}
\email{gezerlis@uoguelph.ca}
\affiliation{Department of Physics, University of Guelph, Guelph, Ontario N1G 2W1, Canada}

\author{Jeremy W. \surname{Holt} }
\email{holt@physics.tamu.edu}
\affiliation{Cyclotron Institute, Texas A\&M University, College Station, TX 77843, USA}
\affiliation{Department of Physics and Astronomy, Texas A\&M University, College Station, TX 77843, USA}

\begin{abstract}
Finding high-quality trial wave functions for quantum Monte Carlo calculations of light nuclei requires a strong intuition for modeling the interparticle correlations as well as large computational resources for exploring the space of variational parameters. 
Moreover, for systems with three-body interactions, the wave function should account for many-body effects beyond simple pairwise correlations. In this work, we design neural networks that efficiently incorporate these factors to generate expressive wave function Ans\"atze for light nuclei using variational Monte Carlo. 
Our neural-network approach for $A=3$ nuclei can capture, already at the level of variational Monte Carlo, the overwhelming majority of the ground-state energy estimated by Green's Function Monte Carlo (GFMC). 
%We can find a 91\% improvement over standard variational Monte Carlo and achieve a ground state energy within $0.45\%$ of the GFMC result for $^3\mathrm{H}$ using the softest chiral interaction with neural networks.
It achieves a ground-state energy within $0.45\%$ of the GFMC result for $^3\mathrm{H}$ using the softest chiral interaction, representing a substantial improvement over standard variational Monte Carlo, which exhibits a $3.7\%$ deviation.
The result indicates the potential of neural networks to construct effective trial wave functions for quantum Monte Carlo calculations.
\end{abstract}

\maketitle
{\it Introduction:}
Over the past few decades, \emph{ab initio} quantum many-body techniques have made remarkable progress in solving the time-independent Schr\"{o}dinger equation for nuclei, leading to a deeper understanding of nuclear structure, dynamics, and the interactions between neutrons and protons \cite{huInitioPredictionsLink2022,lonardoniPropertiesNucleiA162018,lonardoniAuxiliaryFieldDiffusion2018a}. Among the various numerical techniques employed in \emph{ab initio} approaches, the Green’s Function Monte Carlo (GFMC) \cite{lynnQuantumMonteCarlo2017, arriagaThreebodyCorrelationsFewbody1995, pudlinerQuantumMonteCarlo1997, Tews:2015ufa} method is one of the most accurate for calculating the ground-state energies of light nuclei, often serving as a benchmark for other methods.
The GFMC method begins with a carefully chosen trial wave function and then employs imaginary-time evolution to project out the excited-state contributions. 
In this process, the quality of the trial wave function plays a crucial role in controlling the fermion-sign problem, which is necessary to reach dependable answers. 

Conventionally, a trial wave function is obtained using the variational Monte Carlo (VMC) method, where parameterized correlation functions are optimized through standard techniques, such as Powell's method or stochastic gradient descent. These correlation functions are then used to construct the wave function Ansatz. 
For problems involving only two-body interactions, the number of variational parameters typically exceeds 30 \cite{carlsonVariationalMonteCarlo1991} in the standard VMC method, making the optimization of the wave function computationally demanding.
Moreover, one of the more successful VMC frameworks for light nuclei \cite{PhysRevC.32.2105} requires the solution of second-order differential equations \cite{RevModPhys.51.821} from which the wave functions are derived, making the computational burden even heavier. 
Building physically meaningful trial wave functions for complex nuclear interactions requires sophisticated mathematical structures and strong physical intuition (see e.g., Ref.\ \cite{gandolfiAtomicNucleiQuantum2020a} for recent developments in the field).
Therefore, more efficient tools for finding high-quality variational trial wave functions are desired. While VMC is often used to provide wave function inputs to imaginary-time based quantum Monte Carlo (QMC) simulations (e.g., GFMC or auxiliary-field diffusion Monte Carlo),~\cite{lonardoniAuxiliaryFieldDiffusion2018a} the method is also an important standalone tool that can be used to study questions outside the scope of imaginary-time QMC \cite{PhysRevC.83.041001} and that in some cases \cite{PhysRevX.14.021030} can even provide more accurate results than the latter.

According to the Universal Approximation Theorem \cite{cybenkoApproximationSuperpositionsSigmoidal1989, hornikMultilayerFeedforwardNetworks1989a}, neural networks (NN) have the potential to accurately approximate the ground-state wave function of quantum many-particle systems. As a result, numerous studies in both chemistry \cite{pfau2020ferminet, hermannInitioQuantumChemistry2023} and nuclear physics \cite{adamsVariationalMonteCarlo2021, yangDeepneuralnetworkApproachSolving2023, keebleMachineLearningDeuteron2020,Parnes:2025seu} have explored the application of neural networks to the wave function Ansatz in \emph{ab initio} approaches. 
However, in the context of nuclear physics, most of these studies have focused on simplified models of the nuclear force and neglected the complicated tensor, spin-orbit, and three-body terms in microscopic nuclear Hamiltonians. 
% A recent study \cite{yangDeepneuralnetworkApproachSolving2023}, which incorporates neural networks into a modified traditional shell model with manually inputted parameters, has demonstrated the capability to handle tensor and spin-orbit terms, but the complexity arising from pion-exchange interactions remains unaddressed. 
A recent study \cite{yangDeepneuralnetworkApproachSolving2023} encoded nuclear shell structure and introduced spin-isospin-dependent backflow transformation in the neural-network ansatz to address tensor and spin-orbit terms, with parameters trained in a variational way, but the complexity arising from pion-exchange interactions remained unaddressed. 
Moreover, interactions like spin-orbit coupling are not only fundamental in nuclear physics but also in other fields, such as the realization of topological phases in ultracold atoms \cite{PhysRevLett.109.095301,huangExperimentalRealizationTwodimensional2016}.
Thus, we are motivated to develop a neural network architecture capable of handling the full set of interactions that arise from microscopic nuclear two-body and three-body interactions.

In this work, we employ high-precision nuclear forces constructed from chiral effective field theory (EFT) \cite{epelbaum09,machleidtChiralEffectiveField2011},
%, rooted in the fundamental symmetries of QCD , 
which has proven to be highly effective across a broad spectrum of nuclear many-body systems, from finite nuclei \cite{huInitioPredictionsLink2022,PhysRevC.108.064316,lonardoniPropertiesNuclei$A16$2018, chambers-wallMagneticStructure$Aensuremathle10$2024,huInitioComputationsStrongly2024} to infinite nuclear matter \cite{Lim:2025vgb, Shin:2023sei, Drischler:2021kxf, drischlerChiralInteractionsNexttoNexttoNexttoLeading2019, marinoDiagrammaticInitioMethods2024}. 
Importantly, this theory allows for the quantification of uncertainties that arise from (i) the fitting of the free parameters in the theory associated with the short-distance part of the nuclear potential \cite{entem_peripheral_2015,carlsson_uncertainty_2016}, (ii) the choice of resolution scale \cite{Holt:2016pjb,Gasparyan:2022isg,Wen:2023oju} at which nuclear dynamics is resolved, and (iii) the EFT truncation order \cite{Epelbaum:2014efa,furnstahl15_jpg,furnstahl15prc,wesolowski_bayesian_2016,carlsson_uncertainty_2016,Drischler:2017wtt,Drischler:2020yad, PhysRevC.96.024003}. 
%Within this framework, short-range interactions are represented by contact terms, while long-range interactions arise from pion-exchange mechanisms. 
Chiral EFT also naturally incorporates the full complexity of the nuclear force, including two-body tensor and spin-orbit forces as well as three-body forces \cite{gezerlisLocalChiralEffective2014,lynnChiralThreeNucleonInteractions2016, dyhdaloRegulatorArtifactsUniform2016}. 
The complicated features of the full nuclear interaction necessitate the use of highly expressive wave function Ans\"atze for \emph{ab initio} imaginary-time QMC simulations. 

In this Letter, we report that wave functions built from correlation functions generated by neural networks can accurately approximate the ground state of nuclei with mass number $A$ up to 3, incorporating all terms in chiral potentials. %, including tensor, spin-orbit, and three-body interactions without ignoring pion-exchange contributions. 
The designed neural networks take as input both the features of interacting particle pairs as well as the influence from surrounding particles beyond the pairs, enhancing their ability to model systems with three-body forces. 
We train the neural networks using VMC to minimize the estimated ground-state energy and demonstrate that the results closely match the energies obtained from GFMC. 
Additionally, we employ another machine learning technique, the normalizing flow model \cite{Wen:2024shw, Brady:2021plj, PhysRevD.100.034515}, to generate samples for the Monte Carlo estimation of the energy, in order to avoid the correlation length problem that is typically present in classical Markov Chain Monte Carlo algorithms.

{\it Methods:}
A variational trial wave function $\psi$ can be built from the following formula \cite{carlsonQuantumMonteCarlo2015}:
\begin{align}\label{eq:2bpsi}
    | \psi \rangle 
    = \mathcal{S} \prod_{i<j} \Big(1 + \sum_{\rm x} u_{ij}^{({\rm x})} \hat{O}^{({\rm x})}_{ij}\Big) f^{(c)}_{ij} |\Phi\rangle,
\end{align}
where $i$ and $j$ are the indices of the particles, $f^{(c)}_{ij}$ and $\{u^{({\rm x})}_{ij}\}$ are correlation functions that depend on the particle coordinates, and $\mathcal{S}$ is the particle-symmetrization operator. 
The $\{\hat{O}^{({\rm x})}_{ij}\}$ are operators in the set
$
    \{ 
        \bm{\tau}_i \cdot \bm{\tau}_j, 
        \bm{\sigma}_i \cdot \bm{\sigma}_j, 
        (\bm{\sigma}_i \cdot \bm{\sigma}_j) (\bm{\tau}_i \cdot \bm{\tau}_j), 
        S_{ij}, S_{ij} (\bm{\tau}_i \cdot \bm{\tau}_j)
    \}
$, where
$\rm x$ denotes the type of interaction specified by the single-particle spin ($\bm{\sigma}$) and isospin $(\bm{\tau}$) operators, and $S_{ij}$ is the tensor force.
Note that $|\Phi\rangle$ should be an appropriately antisymmetrized quantum state.
For $A=3$ nuclei, $^3{\rm H}$ and $^3{\rm He}$, their antisymmetrized states $|\Phi\rangle$ are written in spin and isospin degree of freedom as 
\begin{align}\label{eq:antsymstate}
    % |\Phi_{^3{\rm H}}\rangle = \mathcal{A} | p_{\uparrow} n_{\downarrow} n_{\uparrow} \rangle, \quad 
    % |\Phi_{^3{\rm He}}\rangle = \mathcal{A} | p_{\uparrow} p_{\downarrow} n_{\uparrow} \rangle,
    |\Phi\rangle = \begin{cases}
    \mathcal{A} | p_{\uparrow} n_{\downarrow} n_{\uparrow} \rangle, & \text{for } {^3{\rm H}},\\
    \mathcal{A} | p_{\uparrow} p_{\downarrow} n_{\uparrow} \rangle,  & \text{for } {^3{\rm He}},
    \end{cases}
\end{align}
where $p(n)$ refers to proton (neutron) and $\uparrow(\downarrow)$ refers to spin up (down), and $\mathcal{A}$ is the antisymmetrization operator. 
For a system with three-body interactions, the trial wave function is usually multiplied by additional operators, 
\begin{align}\label{eq:3bpsi}
    | \psi \rangle \to \Big( 1 + \sum_{i<j<k} \sum_{\rm cyc} \sum_{\rm x} 
    \epsilon^{({\rm x})} \hat{V}^{({\rm x})}_{ijk}\Big) | \psi \rangle.
\end{align}
where $\sum_{\rm cyc}$ runs over the cyclic permutations of a given triple of particle indices $(i, j, k)$, $\epsilon^{({\rm x})}$ is the variational-trainable weight of the corresponding three-body operator $\hat{V}^{({\rm x})}_{ijk}$ with type $\rm x$, independent of the particle permutation $(ijk)$. 
% $\epsilon^{(\mathrm{x})}$ depends only on type-$\rm x$ interaction.
% All $V^{(\mathrm{x})}_{ijk}$ terms with the same interaction type $\mathrm{x}$, regardless of the particle permutation $(ijk)$, share the same weight $\epsilon^{(\mathrm{x})}$.
The three-body operators $\hat{V}^{({\rm x})}_{ijk}$ at N$^2$LO in the chiral expansion are composed of spin-isospin operators which are coupled with the following coordinate space functions 
\begin{align}\label{eq:3bfunction}
    \big\{T^{({\rm x})}_{\alpha\beta}, Y^{({\rm x})}_{\alpha\beta},U^{({\rm x})}_{\alpha\beta}, \delta^{({\rm x})}_{\alpha\beta}\big\}, \quad 
    \alpha \neq \beta = i,j,k.
\end{align}
The expressions for the tensor function \( T^{(\mathrm{x})} \), the Yukawa function \( Y^{(\mathrm{x})} \), and \( U^{(\mathrm{x})} \), as well as their coupling with spin-isospin operators to form three-body interactions, can be found in Ref. \cite{lynnQuantumMonteCarlo2017}. 
% As a part of the original chiral potentials, these spatial functions are identical across all types of three-body interactions. 
In this work, as spatial functions that modify the wave function in Eq.~(\ref{eq:3bpsi}), they do not necessarily have the same functional form as in the original interactions, but we retain how they are coupled with spin-isospin operators.

The Ansatz defined in Eqs.~(\ref{eq:2bpsi}) and (\ref{eq:3bpsi}) is constructed using operators that act on spin-isospin states in conjunction with coordinate-dependent functions. 
The accuracy with which the Ansatz can approximate a target wave function depends on the expressiveness of these coordinate functions. 
We optimize this Ansatz by using neural networks to calculate the correlation functions. 
% Each of them has multiple layers to enhance the expressivity. At each layer, 1-dimensional conventional transformation acts on its input. The activation function RiLU and residual structure are applied to connect layers. 
The neural networks we employ are constructed in two parts, ${\rm NN}_{\rm a}$ and ${\rm NN}_{\rm b}$, where each has multiple layers of 1-dimensional conventional transformations connected by residual structures with the activation function SiLU \cite{elfwingSigmoidWeightedLinearUnits2017}. % forming a residual structure.
For ${\rm NN}_{\rm a}$, the input is the distance $r_{\alpha\beta}$ between the particle pair $(\alpha,\beta)$:
\begin{align}
    \tilde{R}_{\alpha\beta} = {\rm NN}_{\rm a}( r_{\alpha\beta} ), \quad \alpha\neq\beta \leq A,
\end{align}
and its output $\tilde{R}_{\alpha\beta}$ is the feature indicating the influence from surrounding particles.
The second neural network ${\rm NN}_{\rm b}$ generates all of the spatial functions in Eq.~(\ref{eq:2bpsi}) and Eq.~(\ref{eq:3bfunction}):
\begin{align}
    \big\{f^{(c)}_{ij}, u^{({\rm x})}_{ij}, &T^{({\rm x})}_{ij}, Y^{({\rm x})}_{ij},U^{({\rm x})}_{ij}, \delta^{({\rm x})}_{ij}\big\}\nonumber\\
    &=
    {\rm NN_b} \Big( r_{ij} , \sum_{k\neq i,j} \tilde{R}_{ik} + \tilde{R}_{jk} \Big).
\end{align}
With this structure, the correlation functions of the particle pair $(i,j)$ inherently account for the influence of all other third particles $k$, by $\sum_{k\neq i, j} \tilde{R}_{ik} + \tilde{R}_{jk}$, which enhances the performance of the trial wave function for problems with three-body interactions.
Once all of the spatial functions are generated according to the position of particles, one can calculate the trial wave function 
\tcb{(see Supplemental Material for additional details \cite{supplemental})}. 
We note that potentially more sophisticated Ans\"atze for building variational wavefunctions for few-body nuclear systems can be found in Ref.\ \cite{gandolfiAtomicNucleiQuantum2020a}.
% Note that in Eq.~(\ref{eq:3bpsi}), a trainable affine Linear transformation without bias can be applied to combine all the components of $\hat{V}^{(x)}_{ijk}$ so that $\epsilon^{(x)}$ becomes the trainable weights in the linear transformation. With the neural networks ${\rm NN_a}$ and ${\rm NN_b}$, we get the trial wave function $\psi({\bm X}; \bm{\theta})$ of the coordinate of particles ${\bm X}=(\bm{x}_1, \bm{x}_2, \dots, \bm{x}_A)$ and trainable parameters $\bm{\theta}$.

We use VMC to train the neural networks. 
According to the Rayleigh-Ritz principle, the expectation value of the energy in an arbitrary state must be equal to or greater than the ground-state energy. 
With this principle, the loss function is defined as the estimated energy derived from the trial wave function. 
The gradients of the neural networks are then computed in order to optimize the network parameters by minimizing the energy. 
After sufficient training iterations, the neural networks reach a variational wave function with the lowest energy, which should closely approximate the true ground-state wave function. 

\begin{figure}[t]
    \centering
    \includegraphics[width=\linewidth]{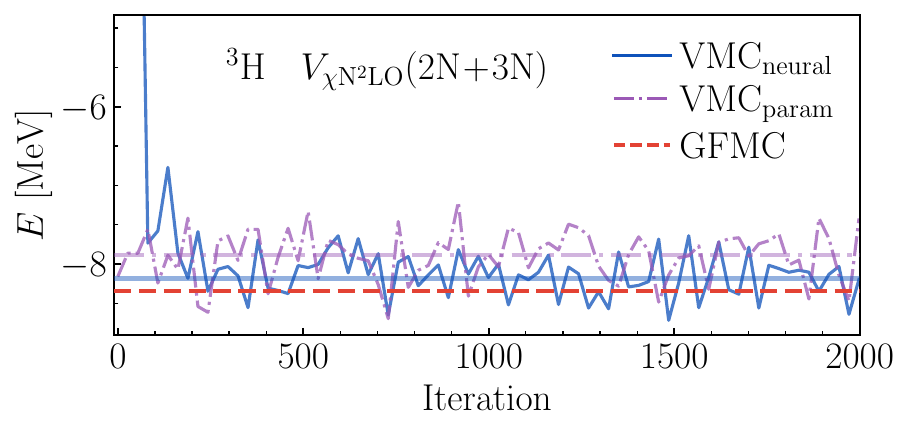}
    \caption{Training of neural networks via VMC for 2,000 iterations, incorporating the full N$^2$LO chiral two- and three-body interactions with $(R_0, \Lambda) = (1.2\ {\rm fm}, 1\ {\rm GeV})$ and using 4,000 samples per iteration. The final results for the neural network VMC (blue horizontal line) and parameterized VMC (purple horizontal dash-dotted line) ground state energies are obtained by averaging over the last 100 iterations and 2,000 iterations, respectively. The benchmark GFMC result (red dashed line) can be found in Ref.~\cite{lynnQuantumMonteCarlo2017}. }
    \label{fig:training}
\end{figure}

To calculate the energy from the wavefunction, we use importance sampling Monte Carlo, which is %most
efficient when samples are drawn from a distribution that matches the target $|\psi|^2$. 
Normalizing flow (nflow) models are employed in this work to generate these samples, avoiding the correlation issues of Metropolis-Hastings and reducing the computational cost.
In normalizing flows, a batch of samples is initially drawn from a simple uniform distribution %, which in the present work is accomplished via a Sobol quasi-random number sequence,
and then passed through a series of change-of-variable transformations with tractible derivatives and Jacobians. 
Through training of the normalizing flow, the resulting distribution can be made to closely resemble the target distribution. 
The energy can then be estimated using $M$ samples of $\bm{X}$ generated by nflow with the following equation 
\begin{align}\label{eq:MCenergy}
    \langle E \rangle &= \frac{\int \prod_{i=1}^A d^{3}\bm{x}_i |\psi(\bm{X};\bm{\theta})|^2 \big(E_{\rm k}(\bm{X}, \bm{\theta}) + V(\bm{X})\big)}{\int \prod_{i=1}^A d^3\bm{x}_i |\psi(\bm{X},\bm{\theta})|^2} \nonumber\\ 
    &\approx \frac{M^{-1} \sum_{m=1}^M |\psi_m(\bm{\theta})|^2 (E_{{\rm k}m}(\bm{\theta}) + V_m)/\rho_m}{ M^{-1} \sum_{m=1}^M |\psi_m(\bm{\theta})|^2/\rho_m},
\end{align}
where $\psi_m(\bm{\theta})$ is the wave function of the $m$th sample $\bm{X}_m$, $\rho_m$ is the corresponding probability distribution function value given by nflow, $E_{\rm k}$ and $V$ are the kinetic energy and potential energy, and $\bm{\theta}$ represents the trainable parameters in the neural networks. 
% Note that the kinetic energy is also a function of trainable parameters ${\bm\theta}$, as it is determined by the derivative of the ${\bm\theta}$-dependent wave function. 
% Because all the coordinate-dependent functions are not directly related to the positions of single particles but the distance of particle pairs, we adopt Jacobi coordinates, so that the kinetic energy contributed from the center of mass, which is not relevant to the ground state of nuclei, is automatically removed in the calculation. 
Since the functions depend on pairwise distances rather than single-particle positions, we adopt Jacobi coordinates to automatically eliminate center-of-mass kinetic energy, which is not relevant for determining the ground states of nuclei.
% In our nflow implementation, Sobol quasi-random sequence is applied when generating uniformly distributed samples to reduce fluctuations of energy estimation during VMC training.  

{\it Results:}
In our experiments, we employ local N$^2$LO chiral interactions \cite{gezerlisLocalChiralEffective2014}. 
For such interactions, the resolution scale is primarily governed by two parameters. 
The first parameter, $\Lambda$, serves as a cutoff in the spectral function regularization of pion-exchange loop diagrams, which prevents the appearance of unnaturally large short-range contributions. 
The second parameter, $R_0$, is used to smear out the Dirac delta function contact terms $\delta(r_{ij})\to \delta_{R_0}(r_{ij})$ in the nucleon-nucleon potential and modify the long-range interactions with a short-distance regulating function according to $V_{\rm long}(r_{ij})\to V_{\rm long}(r_{ij}) (1-e^{-(r_{ij}/R_0)^4})$ \cite{gezerlisLocalChiralEffective2014}. 
% This ensures a smoothening of the impact of the short-distance components of the potential for interparticle distance $r_{ij}$ smaller than $R_0$. 
% For the three-body interactions, 
% the regulator artifacts lead to different types of contact and pion-exchange interactions. In our work, 
In the N$^2$LO chiral potentials, 
we choose the $V_{E\tau}$ and $V_{D2}$ short-distance three-body interactions as in Ref.~\cite{lynnQuantumMonteCarlo2017}. 

\begin{figure}[t]
    \centering
    \includegraphics[width=\linewidth]{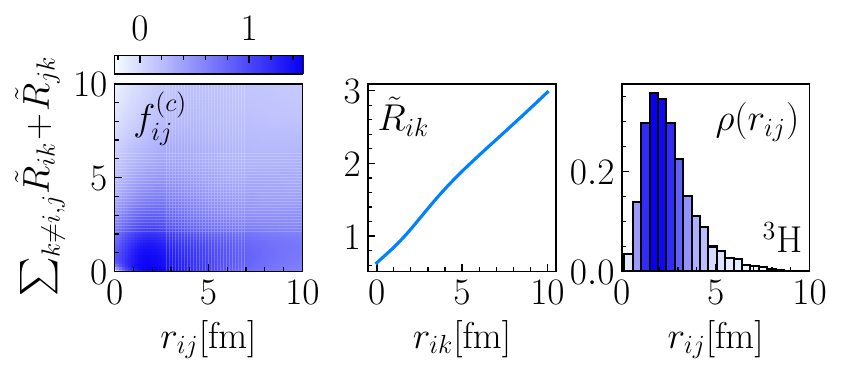}
    \caption{(Left) Correlation function $f^{(c)}_{ij}$ from the well-trained neural networks and (right) the particle pair distance distribution  from nflow model. The well-trained neural networks and nflow are trained for $R_0=1.2$\,fm and $\Lambda = 1.0$\,GeV full chiral potentials as used in Figure \ref{fig:training}. Also shown (middle) is the feature pair distance $\tilde R_{ik}$ output of ${\rm NN_a}$ as a function of the relative coordinate $r_{ik}$.
   }
    \label{fig:correlation_function}
\end{figure}

We begin by training a neural network using the softest N$^2$LO chiral interactions, incorporating both two-body and three-body forces with resolution-scale parameters $(R_0, \Lambda)=(1.2 \text{ fm}, 1 \text{ GeV})$. 
All relevant contributions are included, such as tensor, spin-orbit, charge-independence-breaking (CIB), and charge-symmetry-breaking (CSB) terms without any simplification related to pion-exchange contributions. 
%In VMC, using a physically reasonable initial trial wave function, such as one derived from the Hartree-Fock algorithm or optimized for a similar Hamiltonian, can improve the convergence and efficiency.
We initialize the neural networks by matching their correlation functions to the VMC-optimized two-body correlation functions for the Reid v8 potential \cite{Carlson1991}.
The training process of the neural network (light blue line) is illustrated in Figure \ref{fig:training}, where we observe a rapid decrease in energy, converging efficiently toward the GFMC energy.
We find that compared to the standard parameterized VMC approach~\cite{lonardoniAuxiliaryFieldDiffusion2018a}, whose wave function is used as an input to Auxiliary Field Diffusion Monte Carlo calculations, the wave function from the neural network is a better approximation to the ground state. Specifically, the unparameterized neural network VMC wave function reaches a ground-state energy ($E = -8.2\pm0.3\ {\rm MeV}$) after the training as indicated by the average value over last 100 iterations, compared to the parameterized VMC result 
$E = -7.9 \pm 0.1\ {\rm MeV}$.
A subsequent fine-tuning using a learning-rate scheduler, as detailed in the Supplemental Material \tcb{\cite{supplemental}}, ensures a small learning rate toward the end of training for true convergence. This gives a result of $E=-8.30\pm 0.09 \ {\rm MeV}$ averaged over the last 100 fine-tuning iterations.
%$E = -7.887\pm0.007\ {\rm MeV}$, the accumulated estimation over 1,000 iterations.  
% The softest potential is expected to be the most favorable for the training of the neural networks, as it yields small uncertainties in the estimated energy. 
% More precise energy estimation can reduce the gradient noise of the trainable parameter $\bm{\theta}$, clarifies optimization direction and improves neural network training efficienty. 
% A well-trained model for this interaction can be viewed as a pre-trained model for other interactions, facilitating their training. 

\begin{figure}[t]
    \centering
    \includegraphics[width=\linewidth]{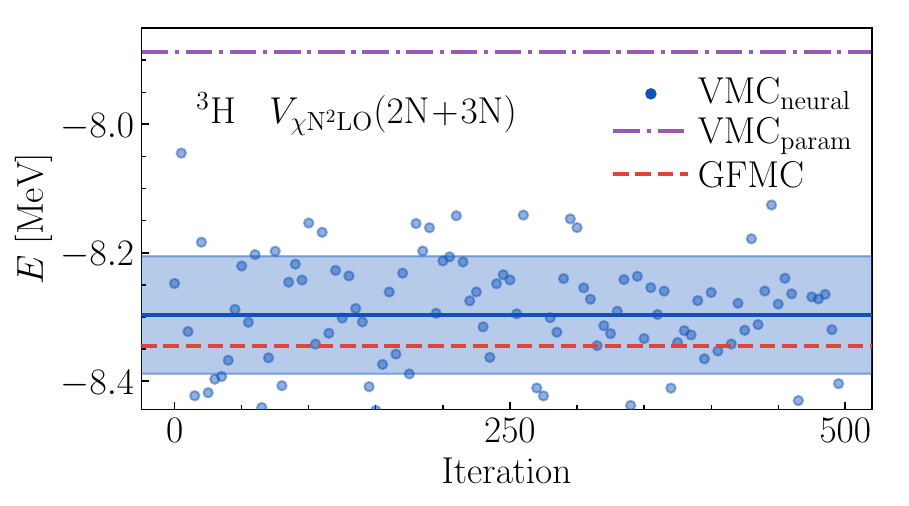}
    \caption{Evaluation of the $^3{\rm H}$ ground-state energy using well-trained neural networks with the full version of the N$^2$LO chiral interaction with $(R_0,\Lambda)=(1.2\ {\rm fm}, 1\ {\rm GeV})$. The average energy (blue horizontal line) and standard deviation (blue-shaded region) from neural networks are estimated over 500 iterations with 40,000 samples per iteration. The standard VMC result (purple line) is the same as in Figure \ref{fig:training}. }
    \label{fig:eval}
\end{figure}

The correlation functions for an interacting particle pair are influenced not only by $r_{ij}$ but also by other surrounding particles, where the impact from the latter are captured by ${\rm NN_b}$ and represented as $\sum_{k\neq i,j} \tilde{R}_{ik}+\tilde{R}_{jk}$. 
%In our work, this surrounding-particle dependence is captured by ${\rm NN_b}$ and represented as $\sum_{k\neq i,j} \tilde{R}_{ik}+\tilde{R}_{jk}$. 
The middle panel of Figure \ref{fig:correlation_function} illustrates how this feature is encoded through $\tilde{R}_{ik} = {\rm NN_a}(r_{ik})$, which gradually increases close to a linear trend.
Regarding the correlation function $f^{(c)}_{ij}$ shown in the left panel of Figure \ref{fig:correlation_function}, one sees that it starts from a small value at $r_{ij}=0$, rapidly increases, and then gradually decreases as the interparticle distance grows. 
%$f^{(c)}_{ij}$ evolves smoothly and uniformly along both input dimensions $r_{ij}$ and $\sum_{k\neq i, j}\tilde{R}_{ik} + \tilde{R}_{kj}$, indicating ${\rm NN_a}$ is actually a preprocessing layer that rescale and shift the inputs of surrounding particle distances, ensuring the two inputs for ${\rm NN_b}$ are of the same magnitude.
Since $f^{(c)}_{ij}$ primarily governs the magnitude of the wave function, its suppression at short $r_{ij}$ implies that the probability of finding two particles closely separated is extremely low, which is consistent with the repulsive nature of short-range nuclear interactions. 
Moreover, $f^{(c)}_{ij}$ exhibits a strong dependence on third-particle effects, particularly when the two interacting particles are closely surrounded by others. 
This is reflected in the large magnitude of $f^{(c)}_{ij}$ when $\sum_{k\neq i,j}\tilde{R}_{ik} + \tilde{R}_{jk}$ is small. 
% However, as surrounding particles move away, the influence from many-body features diminishes.
The right panel shows the $r_{ij}$ distribution from nflow. 
It has features similar to those observed in the correlation functions and therefore is suitable
%to generate samples 
for the importance sampling Monte Carlo estimation of the energy. % in Eq.~(\ref{eq:MCenergy}).

 % accumulated gradient techniques and a gentle learning rate. 
% Once a model is well-trained, it is used to evaluate the energy. 
We next use the well-trained neural network models to derive precise evaluations of the ground state energy, using a gentle learning rate for further updating.
In Figure \ref{fig:eval}, we perform 500 iterations to estimate the energy from the neural-network-generated wave function with nflow importance sampling.  
%The estimation is carried out using importance sampling Monte Carlo, where both the samples and their corresponding probabilities are provided by the nflow. 
% We observe that all energy values throughout the iterations remain within the one-sigma region of the GFMC results, demonstrating that the nflow is well-suited for importance sampling Monte Carlo, as it yields stable and high-precision results. 
% The average energy estimation (blue line in the left panel) and the accumulated estimation (blue line in the right panel) across all iterations closely align with the GFMC result, with only a $0.45\%$ deviation above the GFMC result. 
The average energy from the neural network model is $E = -8.30 \pm 0.09\ {\rm MeV}$, which closely matches the GFMC result $E=-8.34\ {\rm MeV}$, falling within one-half of a standard deviation and differing by only $0.45\%$.
The uncertainty of $0.09\ {\rm MeV}$, defined as the standard deviation over the 500 estimates, indicating the small fluctuation in the energy estimation with nflow.
This indicates that the neural-network-generated wave function can effectively approximate the ground state of $^3{\rm H}$ with the soft-core interaction. %without any imaginary-time evolution, even in the presence of complicated pion-exchange interactions, as well as tensor, spin-orbit, CIB, and CSB interactions. 
%Compared with the standard parameterized VMC approach (purple line), we can find a 91.2\% improvement, estimated by $(E_{\rm param} - E_{\rm neural})/(E_{\rm param} - E_{\rm GFMC})$, from neural network VMC. 
%Moreover, the relative uncertainty for the 500 iterations is $1\%$, which encompasses the GFMC result, indicating that the nflow estimation performs reliably and is suitable for stable neural network training. 

\begin{figure}[t]
    \centering
    \includegraphics[width=\linewidth]{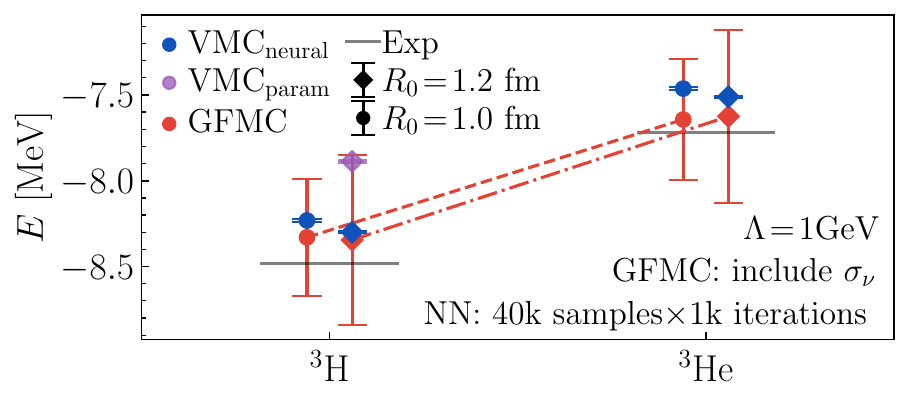}
    \caption{Ground-state energies for $A=3$ nuclei obtained from well-trained neural networks with Coulomb and full chiral potentials. The uncertainties shown for the GFMC calculations also include contributions from chiral EFT truncation errors (not included in the VMC$_{\rm neural}$ error bands). The standard VMC result is from the same estimation in Figure \ref{fig:training}.
    The experimental data (gray horizontal lines) can be found in Refs.~\cite{lynnQuantumMonteCarlo2014, ParticleDataGroup:2024cfk}.
    %Complex pion-exchange interactions, as well as tensor, spin-orbit, CIB, and CSB interactions, are considered in all the calculations. 
    }
    \label{fig:nuclei}
\end{figure}

\begin{table}[b]
\centering
    %\begin{tabular}{|c|c|c|c|}
    %\hline
    %    \multicolumn{4}{|c|}{ 
    %        \begingroup
    %        \boldmath
    %        ${^3}{\rm H}$: $E = E_{\rm k} + V_{\rm 2N}$ 
    %        \endgroup
    %    } \\ \hline
    %    $R_0$ [fm] & $E_{\rm NN}$ [MeV] & $E_{\rm GFMC}$ [MeV] & $|\Delta E|/|E_{\rm GFMC}|$ \\ \hline
    %    1.0 & $-7.338 \pm 0.008$ & $-7.554 \pm 0.007$ & 2.8\% \\ \hline
    %    1.1 & $-7.500 \pm 0.006$ & $-7.625 \pm 0.005$ & 1.6\% \\ \hline
    %    1.2 & $-7.678 \pm 0.005$ & $-7.740 \pm0.005$ & $\bm{0.8\%}$ \\ \hline
    %\end{tabular}
\begin{tabular}{|ccccc|}
\hline
\multicolumn{5}{|c|}{ 
    \begingroup
    \boldmath 
    $E = E_{\rm k} + V_{\chi {\rm N}^2 {\rm LO}}({\rm 2N})$ 
    \endgroup
    }  \\ \hline
\multicolumn{1}{|c|}{} 
& \multicolumn{1}{c|}{$R_0$ {[}fm{]}} 
& \multicolumn{1}{c|}{$E_{\rm neural}$ {[}MeV{]}}  
& \multicolumn{1}{c|}{$E_{\rm GFMC}$ {[}MeV{]}} 
& $|\Delta E|/|E_{\rm GFMC}|$ \\ \hline
\multicolumn{1}{|c|}{\multirow{3}{*}{$^3{\rm H}$}} 
& \multicolumn{1}{c|}{$1.0$} 
& \multicolumn{1}{c|}{$-7.338 \! \pm \! 0.008$} 
& \multicolumn{1}{c|}{$-7.554 \! \pm \! 0.007$} 
& $2.9\% $ \\ \cline{2-5} 
\multicolumn{1}{|c|}{} 
& \multicolumn{1}{c|}{$1.1$} 
& \multicolumn{1}{c|}{$-7.500 \! \pm \! 0.006$} 
& \multicolumn{1}{c|}{$-7.625 \! \pm \! 0.005$} 
& $1.6\%$ \\ \cline{2-5} 
\multicolumn{1}{|c|}{}  
& \multicolumn{1}{c|}{$1.2$} 
& \multicolumn{1}{c|}{$-7.678 \! \pm \! 0.005$} 
& \multicolumn{1}{c|}{$-7.740 \! \pm \! 0.005$} 
& $0.8\%$ \\ \hline
\multicolumn{1}{|c|}{\multirow{2}{*}{$^2{\rm H}$}}  
& \multicolumn{1}{c|}{$1.0$} 
& \multicolumn{1}{c|}{$-2.217\!\pm\!0.005$} 
& \multicolumn{1}{c|}{$-2.21 \! \pm \! 0.02$} 
& 0.3\% \\ \cline{2-5} 
\multicolumn{1}{|c|}{} 
& \multicolumn{1}{c|}{$1.2$} 
& \multicolumn{1}{c|}{$-2.212\!\pm\!0.004$} 
& \multicolumn{1}{c|}{$-2.20 \! \pm \! 0.03$} 
& 0.5\% \\ \hline
\end{tabular}
\caption{$^{3}{\rm H}$ and $^{2}{\rm H}$ ground state energies obtained from ${\rm NN_b}$-only neural networks compared to GFMC. The neural network results are the weighted estimation over 1,000 interactions with 40,000 nflow-generated samples per iteration.
The energy difference is defined as $\Delta E = E_{\rm NN} - E_{\rm GFMC}$. 
The GFMC results are taken from Ref.~\cite{lynnQuantumMonteCarlo2014} for $^3{\rm H}$ and Ref.~\cite{lynnQuantumMonteCarlo2017} for $^2{\rm H}$.}
\label{tab:2bH2}
\end{table}

We next evaluate the ability of neural network VMC to model nuclear few-body wavefunctions resulting from finer-resolution chiral potentials, which generally have a stronger short-distance repulsive core. Specifically, we employ the $R_0=1.0$\,fm N$^2$LO chiral nuclear force, including also Coulomb interactions. 
% The Coulomb interactions play a crucial role in the case of $^3{\rm He}$. 
%To achieve precise energy estimation, % from neural networks, 
%we use 40,000 samples per iteration and perform 1,000 iterations. 
The results for the $^3$H and $^3$He ground-state energies, each obtained from 1,000 iterations with 40,000 samples per iteration, are presented in Figure \ref{fig:nuclei}.
We find that all neural network VMC predictions fall within the one-sigma GFMC error band from Ref.\ \cite{lynnQuantumMonteCarlo2017}, which we note includes an estimate of uncertainties coming from the truncation of the chiral expansion (not included in our calculations). We see that
neural network VMC performs particularly well when dealing with soft interactions ($R_0 = 1.2 \text{ fm}$), likely due to a more accurate and stable energy estimation during training. Additionally, the discrepancy between neural network VMC results and the central values for GFMC for $^3{\rm H}$ is smaller than for $^3{\rm He}$ when the same types of chiral potentials are considered. 
This may reflect the limitations of the wave function Ansatz, where isospin symmetry is preserved, whereas 
% in the case of $^{3}{\rm He}$, 
the Coulomb interaction breaks this symmetry in $^{3}{\rm He}$. We also show for comparison the $^3$H ground-state energy employing the $R_0 = 1.2$\,fm N$^2$LO chiral potential using the traditional parameterized VMC method (shown as the purple diamond). 
Compared with the energy estimation for $^3{\rm H}$ using the $R_0=1.2\ {\rm fm}$ chiral interaction in Fig.~\ref{fig:eval}, the accumulated estimation from 1,000 iterations yields a smaller uncertainty, consistent with statistical expectations.

In order to further benchmark the neural networks for generating nuclear wave function correlations, we also analyze $^2$H and $^3$H without three-body interactions. %In this case, the influence of a third-particle associated with $\tilde{R}_{ik}+\tilde{R}_{jk}$ can be neglected. When considering only two-body interactions, influence from third particles $\tilde{R}_{ik}+\tilde{R}_{jk}$ beyond particle pairs $(i,j)$ can be neglected. Consequently, ${\rm NN_a}$ becomes unnecessary, and the correlation functions can be computed solely using ${\rm NN_b}$ via $\{f_{ij}^{(c)}, u_{ij}^{(x)}\} = {\rm NN_b}(r_{ij})$. The trial wave function is then constructed according to Eq.~(\ref{eq:2bpsi}). 
In Table \ref{tab:2bH2}, we show the calculated ground-state energies of $^{2}{\rm H}$ and ${^3}{\rm H}$ using different N$^2$LO two-body-only chiral potentials and compare the results with those from GFMC \cite{lynnQuantumMonteCarlo2014,lynnQuantumMonteCarlo2017}. 
Once again, the softest interaction $(R_0 = 1.2 \text{ fm})$ is favored for the three-body system $^3{\rm H}$, exhibiting only a 0.8\% deviation from the GFMC result. 
For $^2\rm{H}$, the neural network results are nearly indistinguishable from the GFMC results within a one-sigma uncertainty.
%This demonstrates that ${\rm NN_b}$-only neural networks are well-suited at modeling wave functions for complicated two-body interactions. 

{\it Summary and Outlook:}
In this Letter, we have designed neural networks capable of generating more expressive wave functions for VMC studies of light nuclei, including both two-body and many-body correlations. %The neural network framework consists of two components: ${\rm NN_a}$ and ${\rm NN_b}$, where ${\rm NN_a}$ captures features from particles beyond the interacting particle pair, while ${\rm NN_b}$ takes as input both the interacting particle pair distance and the many-body features extracted by ${\rm NN_a}$ to compute the correlation functions. 
%These correlation functions, combined with spin-isospin operators, act on antisymmetrized quantum states to produce trial wave functions for light nuclei. 
%With this structure, many-body features are encoded to make sure neural networks can deal with \emph{ab initio} problems containing three-body interactions. 
% The neural networks are trained via VMC to approximate the nuclear ground state, where the energy is estimated by importance sampling Monte Carlo equipped with normalizing flow neural network models. % by minimizing the energy. 
% In the VMC framework, an nflow model is employed, with its target distribution corresponding to the trial wave function, 
% to generate samples efficiently 
% for importance sampling Monte Carlo during energy estimation.
We find that employing local chiral interactions at N$^2$LO, which include complicated pion-exchange interactions, as well as tensor, spin-orbit, charge-independence breaking, charge-symmetry breaking, and three-body interactions, the neural-network VMC method produces ground-state energies that are very close to those computed using GFMC. Especially with the soft-core interactions, the energy obtained for ${^3}{\rm H}$ using neural networks deviates by only 0.45\% from the GFMC energy. %, leading to a \tout{91.2\%} improvement over 
%compared to standard parameterized VMC, which exhibits a $3.7\%$ deviation.

%Additionally, Coulomb interactions are explicitly considered in ${}^3\mathrm{He}$. 

%To assess the quality of the neural network-generated wave functions, we compare the resulting energy estimates with those obtained from GFMC. 
%The results demonstrate that the neural-network-based wave functions closely approximate the nuclear ground state, yielding energies that are very close to those computed using GFMC already at the VMC level. 

% Furthermore, using soft-core interactions ensures stable and accurate energy estimations, facilitating effective optimization of the neural networks and improved wave function approximation. 
%As a result, the energy obtained for ${^3}{\rm H}$ using neural networks deviates by only 0.45\% from the GFMC energy.
% Compared with the standard VMC, the improvement is 91.2\% from neural networks.
%The strong dependence of the correlation functions on the encoded many-body features revealed by ${\rm NN_a}$ reveals that many-body features play a crucial role in determining the wave functions. 
%The stable and precise estimation of energies indicates that nflow is able to generate samples whose distribution follows a trial wave function, thereby enhancing the efficiency of neural network training within the VMC framework.

Currently, the framework incorporates six types of correlation functions in Eq.~(\ref{eq:2bpsi}). % However, numerous studies suggest that incorporating additional correlation functions coupled with other spin-isospin operators could be beneficial, warranting further investigation.
Additional correlation functions coupled with other spin-isospin operators are worth investigating. In this work, the antisymmetrization is performed in spin-isospin space, which limits the Ansatz to $A\leq 4$ nuclei. However, it can be extended to treat a larger range of nuclei by utilizing, e.g., neural networks to construct Slater determinants for the spatial orbitals of individual particles following the structure proposed by Ref.~\cite{pfau2020ferminet} and combining them with neural-network correlation functions developed in this work.

%The normalizing flow exhibits strong performance for importance sampling in Jacobi coordinates, highlighting its potential to generate samples for Monte Carlo calculations in heavier systems without explicitly employing Jacobi coordinates, such as sampling individual particles in no-core shell-model configurations. 

%Given the high quality of the neural-network wave functions, they can serve as initial trial wave functions for other Quantum Monte Carlo (QMC) methods, such as GFMC or auxiliary-field diffusion Monte Carlo, enabling further optimization, potentially enhance both the efficiency and accuracy of QMC calculations.
% This approach has the potential to enhance both the efficiency and accuracy of QMC calculations.
%In this work, the antisymmetrization is performed in the spin-isospin space, which limits the ansatz to $A\leq 4$ nuclei. However, it can be extended to the full space of a fermion system, incorporating spin, isospin, and coordinate degrees of freedom. This can be achieved by utilizing neural networks to construct Slater determinants, where the orbitals are defined by both spin-isospin and coordinate variables. The integration of NN-based correlation functions with NN-generated Slater determinants presents a promising approach for \emph{ab initio} calculations of heavier nuclei.

\begin{acknowledgments}
A.G. would like to acknowledge discussions with Ryan Curry. J.W.H.\ thanks Miski Nopo for insights during the initial stages of the project.
The work of J.W.H. was supported
by the National Science Foundation under Grant No.\ PHY2209318. 
The work of A.G. was supported by the Natural Sciences and
Engineering Research Council (NSERC) of Canada and
the Canada Foundation for Innovation (CFI).
The work of P.W. was supported by 
Texas A\&M Nuclear Solutions Institute, 
Cyclotron Institute.
Portions of this research
were conducted with the advanced computing resources
provided by Texas A\&M High Performance Research
Computing, Compute Ontario through the Digital Research Alliance of Canada, and by the National Energy Research Scientific Computing Center (NERSC), which is
supported by the U.S. Department of Energy, Office of
Science, under contract No. DE-AC02-05CH11231
\end{acknowledgments}

%\bibliography{cite}
\bibliographystyle{apsrev4-2}

\end{document}